\documentclass{article}
\usepackage{wds10,epsf}

\usepackage[cp1251]{inputenc}
\usepackage[T2A]{fontenc}
\usepackage[russian]{babel}
\usepackage{amsmath}

\lefthead{SEROV, KOVAL}
\righthead{RSRG Approach}

\begin{document}

\title{Real-Space Renormalization Group (RSRG) Approach to Quantum Spin Lattice Systems}

\author{A.\,S. Serov$^{a}$, G.\,V. Koval$^{b}$}

\affil{M.\,V. Lomonosov Moscow State University, Faculty of Physics, Moscow, Russia.}

E-mail: $^a$a.s.serov@gmail.com,
$^b$gkoval@phys.msu.ru

\begin{abstract}
The paper describes an explicit variational modification of the standard RSRG method and its application to quantum spin lattice systems. The modified approach is applied to exactly solvable ITF, $XX$ and isotropic Heisenberg models. Better upper bounds for the systems' ground state energy are obtained as compared to the standard RSRG approach.

\end{abstract}


\section{Introduction}

Renormalization group is a method of studying systems with a large number of strongly correlated degrees of freedom. The main idea this method makes use of, is to decrease the number of degrees of freedom of the system, so as to preserve the information about the essential physical properties of the system and omit those aspects, which are not important for the considered phenomena.

Application of the renormalization group in real space to lattice systems implies construction of a new smaller system corresponding to the initial one with new interactions between the degrees of freedom. In the new system there are less degrees of freedom, and the interactions between them are by some rules expressed through the interactions in the initial system. One of the tasks of renormalization group is to obtain the rules, which define the transformation of "old"\, interactions into "new"\, ones. Analyzing such equations one can determine qualitatively the structure of the phase diagram of the studied system, approximately locate the critical points and obtain the critical exponents. It is also possible to get such quantitative characteristic of the systems as an upper bound for the ground state energy.

Still the idea of renormalization group can be applied differently to the same system, so that the transformations of "old"\, interactions into "new"\, ones will have different forms. The difference and the freedom here lies in choosing the state space of the new system and how exactly the states of the initial system will be projected onto the state space of the new system.

One of the possible applications of RSRG method is to study lattice models. Lattice models can describe, for instance, such physical systems as simple ferromagnetics, antiferromagnetics and ferrites [Cardy, 1996]. In these cases spin $s=1/2$ is placed in the sites of the lattice. Furthermore, such models can describe magnetic substance with crystalline lattice with vacancies or simple liquid. Thus, application of the RSRG ideas to the latter system makes it possible to analyze phase transitions in liquid-gas systems.

RSRG is only one of the possible methods that can be chosen to study lattice systems with a large number of strongly correlated degrees of freedom. Alternative to it are, for example, Bethe ansatz, the mean field approximation (MFA) and Density Matrix Renormalization Group method (DMRG,~[White,~1992]). See~[Cardy, 1996] and~[Mart\'in-Delgado, 1996] for a review on other applicable methods. Still in application to some problems these methods prove to have some drawbacks. Thus, Bethe ansatz gives exact theoretical result for only a few 1D lattice models and fails in 2D. The MFA doesn't take into consideration any fluctuations in the systems, and that makes it not suitable for study of critical phenomena (phase transitions, for instance). As to DMRG method, it produces extremely accurate results for the ground state and excitations energy if applied to one-dimensional systems, but it turns out that it's hard to apply it to the problems with more than one dimension, though some progress in this field has been already achieved.

The current paper is devoted to how the accuracy of the results can be systematically improved compared to standard RSRG method [Burkhardt et al, 1982] by the use of variational ideas and symmetry considerations. 

The main difference between the method proposed in the article and the standard RSRG method and many its popular modifications (see for example, the idea of superblocks in [Langari, Karimipor, 1998] and in [Malrieu et al, 2001]) lies in how we choose new spin states space. Most of the methods diagonalize the hamiltonian of a block or a superblock on some step or even on every step, which can be very time-demanding. We propose a modification that doesn't use diagonalization at all. Most methods include some additional considerations that somehow choose which of the states of the diagonalized matrix should be kept and would represent a hamiltonian after a renormalization step. In our approach variation of the ground state energy is the thing that chooses new states, no additional considerations required. Somewhat similar idea was proposed in the paper [Drell et al., 1977], although the authors didn't manage to find a closed equation for the ground state energy and had to vary a parameter after a fixed number of iterations, whereas we vary it after an "infinite"\, number of iterations.

S.~\"Ostlund and S.~Rommer [\"Ostlund et al.,~1995] have showed that DMRG method in 1D can be formulated variationally with a wave function chosen in a matrix product form, although the variational procedure we propose is different from the one these authors use.

Another feature of our modification of the RSRG method lies in that it always gives an estimate from above for the ground state energy, whereas other modifications sometimes give just an estimate for the true value not knowing whether the real value is greater or lower than the acquired result. We in the meantime always know that the value we get is greater than the true value, and we also have some ideas on how to devise a method that would give an estimate from below for the value. Having these two values we would always be able to provide an interval in which the true value lies with 100\% accuracy.

\section{Standard RSRG method}

To describe the new ideas that we develop, we should first discuss the realization of the standard RSRG method [Drell et al, 1977; Burkhardt et al, 1982].

We will consider lattice systems with spin $1/2$ in their sites, taking 1D spin chain as an example. Let's then choose the hamiltonian of quantum Ising model in transverse field (ITF model) as interaction between the spins of the lattice:

\begin{equation}
H=-J\sum_{<i,j>}S_i^xS_j^x-h\sum_iS_i^z,
\label{itf}
\end{equation}
where $S_i^x, S_i^z$ are the $x$- and $z$-components of the spin $1/2$ at the $i$-th site of the lattice respectively; constant $J$ characterizes the intensity of spin-spin interaction, and the $h$ constant determines the intensity of spin interaction with external magnetic field; index $<i,j>$ in the sum means summing over the pairs of nearest neighbours.

The described conceptions can be nevertheless applied to other interactions, for example to interactions in $XX$-model or isotropic Heisenberg model.

Realization of the renormalization group idea in real space is based on Kadanoff block-spin transformation [Kadanoff, 1966]. This transformation splits the initial lattice into non-crossing equally sized blocks of spins, and all the blocks together fully cover the lattice. Then a new smaller lattice of effective spins is created, in which each spin corresponds to a block of the initial lattice. For the reasons of simplicity for our consideration we take two spins in a block.

Then, following the RSRG ideas, we split the hamiltonian~(\ref{itf}) into intrablock~(B) and interblock~(BB) parts:

$$
H=\sum_{j=1}^{N/2}\left(H_{j,B}+H_{j,BB}\right)
$$
\begin{equation}
\label{HB}
H_{j,B}=-JS_{2j-1}^xS_{2j}^x-h\left(S_{2j-1}^z+S_{2j}^z\right)
\end{equation}
$$
H_{j,BB}=-JS_{2j}^xS_{2j+1}^x
$$

Each block contains two sites and, therefore, represents $2^2$ degrees of freedom and a state space with the same dimension. According to Kadanoff transformation, in order to set an effective spin $1/2$ corresponding to this block, it's necessary to select only two states from the mentioned state space. As we are planning to find the whole system's ground state, it's possible to suppose that the system's ground state is built of the block states with the lowest energies. That means, we should diagonalize one-block hamiltonian and select two eigenstates with the lowest eigenvalues as the states of a new effective spin.

Let's denote selected states of the new effective spin as $|\widetilde{\downarrow}>$ and $|\widetilde{\uparrow}>.$

If now we project the intrablock hamiltonian of the $j$-th block on this state space, we obtain the hamiltonian of effective spin interaction with an external field

$$
\widetilde{H}_{j}=\varepsilon_1|\widetilde{\uparrow}><\widetilde{\uparrow}|+
\varepsilon_2|\widetilde{\downarrow}><\widetilde{\downarrow}|=
-\widetilde{h}\widetilde{S}_j^z+\widetilde{C},
$$
where $\varepsilon_1$ and $\varepsilon_2$ are the lowest eigenvalues of the intrablock hamiltonian $H_B$, $\widetilde{h}$ and $\widetilde{C}$ are renormalized external field and a new constant respectively, $\widetilde{S}_j$ is the effective spin operator.

Projecting the interblock hamiltonian we get a part of a hamiltonian, which describes the interaction of the new spin with its nearest neighbours:

$$
\widetilde{H}_{j,int}=-\widetilde{J}\widetilde{S}_j^x\widetilde{S}_{j+1}^x,
$$
where $\widetilde{J}$ is the renormalized spin-spin interaction.

So the hamiltonian of the initial system under this procedure has transformed into a hamiltonian of the same form, in which the values of interactions are the functions of the initial interactions; there also appeared a constant summand to the system energy, and the number of the degrees of freedom has decreased by $2^{N/2}$.

Then the standard approach analyzes the transformation rules $\widetilde{J}=\widetilde{J}(J,h), \widetilde{h}=\widetilde{h}(J,h)$ of "old"\, interactions into "new"\, ones, and from this analysis it makes conclusions about the system's phase diagram, finds the critical points and critical exponents. By iterating the mentioned procedure many times, the ground state energy is numerically found [Drell et al, 1977; Burkhardt et al, 1982].

\section{Modification to the standard approach improving the estimate for the ground state energy}

We show later, that the described RSRG method is imperfect in the part of getting an estimate for the system's ground state energy, what has been also mentioned by other authors [Drell et al, 1977]. That's why this paper suggests some ideas, which help to obtain more accurate results.

\subsection{The structure of new spins states}

To improve the accuracy of the RSRG method, it's necessary to alter the structure of the new effective spins states. Namely, for the new effective "down"\, spin state we suggest taking a linear combination with arbitrary parameters of all basis vectors of the initial block, containing even number of excitations; for the new "down"\, spin state --- a linear combination of vectors with odd number of excitations. The term excitation (magnon) here denotes a spin directed "up"\, in the initial basis vector.

Let's explain it on the example of two spins in a block. We choose the basis of the initial state space of two spins in the form of $|\downarrow\downarrow>,|\uparrow\downarrow>,|\downarrow\uparrow>,|\uparrow\uparrow>.$ According to what was said, the "up"\, and "down"\, states of the effective spin should be chosen as:
\begin{align}
\label{downState}
|\widetilde{\downarrow}>=\frac{|\downarrow\downarrow>+\alpha_1|\uparrow\uparrow>}
{\sqrt{1+\alpha_1^2}},\\
\label{upState}
|\widetilde{\uparrow}>=\frac{|\downarrow\uparrow>+\alpha_2|\uparrow\downarrow>}
{\sqrt{1+\alpha_2^2}},
\end{align}
where $\alpha_1,\alpha_2$ are some arbitrary parameters.

Let's now notice that the initial block possesses a symmetry of reflection against its center. To preserve this symmetry we impose a corresponding additional condition on the new states. Taking this into account, we rewrite the states~(\ref{downState}),(\ref{upState}) in the following form:
\begin{align}
\label{symDownState}
&|\widetilde{\downarrow}>=\frac{|\downarrow\downarrow>+\alpha_1|\uparrow\uparrow>}
{\sqrt{1+\alpha_1^2}}\\
\label{symUpState}
&|\widetilde{\uparrow}>=\frac{|\downarrow\uparrow>+|\uparrow\downarrow>}
{\sqrt{2}}
\end{align}

The idea of such alteration of the structure lies in that the described above supposition, that a system's ground state is built of the ground states of each block, is not enough accurate. Really, a system's ground state is a superposition of all the basis states, and omission of high-energetic block states leads to errors [see Mart\'in-Delgado, 1996]. Thus, we suppose, that the "right"\, choice of introduced parameters will improve the accuracy of the results we get.

For the first time the idea of choosing the new spin state space as a linear combination with arbitrary parameters was introduced in the paper [Drell et al, 1977] for ITF model, though it was developed in another way there. The authors of that article didn't get a closed equation for the system's ground state energy. As a result they made some guesses on the form of the parameters dependency on the interaction values and used them in their optimization. Below we show, how a closed expression for the ground state energy can be obtained, which then can be minimized against the parameters without any additional suppositions. It will also allow us to introduce different parameters on different steps of renormalization, and to minimize in the end the energy against all the parameters at once.

We would like to mention also, that the possibility of minimization of the estimate for the energy obtained with the help of RSRG method rests upon the fact, that the decrease of the number of degrees of freedom and the reduction of the state space of the initial system is always performed so, that every previous state space includes every next one as a subspace. Therefore, a vector that minimizes the energy in the new state space is also a member of the initial state space of the system under consideration. In other words, the system energy obtained after the minimization is always an upper bound for the ground state energy of the initial system.

\subsection{Sets of parameters}

The obtained estimate for the energy can be further improved if, as mentioned above, we introduce different sets of parameters for every step of renormalization. A little disadvantage of such approach lies in that with an infinite number of parameters one cannot get a closed (finite) expression for the system's energy, and so cannot minimize it. Still this disadvantage can be mended if different parameters are introduced only on several first steps and starting from some step all the parameters are taken to be the same. Such approach makes it possible to get a closed expression for the energy, which can be optimized in practice.

We will also mention that the analysis of the parameters values obtained during the optimization process can shed light on the ground state structure which is not always clear.

\subsection{Multi-steps blocking}

At last, a multi-steps blocking can be used in order to improve the accuracy of the estimates. For example, for a block of eight spins one can set one corresponding effective spin in the following way: first to group the spins into blocks of four spins and to conduct the described RSRG transformation for them, and then to consider the group of two obtained spins. We should stress, that the difference of such approach and a simple periodic change of the block size lies in the fact that only the initial block of eight sites is supposed to possess a reflection symmetry against its center, and the intermediate blocks of four and two spins are not supposed to have such symmetry of their own. Such an algorithm makes it possible to conduct "indirect"\, calculations for large clusters, for which the "direct"\, implementation of the procedure described above becomes complicated.

Generally speaking, the order of selecting the sizes of blocks to approximate a big one also matters. It means, that for a cluster of eight sites grouping the spins by four and then by two, and alternatively, by two and then by four can yield different results.
The comparison of these results --- why one order of choosing blocks sizes gives better results, while the alternative gives worse --- can give soil for investigation of the system's ground state structure.

\section{An example of application of the modified RSRG method to ITF model}

Let's now consider the application of the mentioned ideas to modify the standard RSRG method described before.

Again, we will divide the sites of the initial lattice into blocks of two sites and will split the system hamiltonian into intrablock and interblock parts. However, now we don't need to diagonalize the intrablock part, we simply choose the new spins states as~(\ref{symDownState}) and~(\ref{symUpState}). After that the intra- and inter-block parts of the initial system hamiltonian are projected to the new system's state space, built of the sets of~(\ref{symDownState}) and~(\ref{symUpState}) states for each block.

Calculations show, that the hamiltonian of a new effective system, as in the standard approach, will have the following form:
$$
\widetilde{H}=\sum_{j=1}^{N/2}\left(-\widetilde{J}\widetilde{S}_j^x\widetilde{S}_{j+1}^x-
\widetilde{h}\widetilde{S}_j^z+\widetilde{C}\right)\!,
$$
but where $\widetilde{J}=\widetilde{J}(J,h;\alpha_1)$ and $\widetilde{h}=\widetilde{h}(J,h;\alpha_1)$ now depend on $\alpha_1$ parameter.

In order to get a closed expression for the ground state energy depending on parameters, it's useful to rewrite the transformation of interactions $J$ and $h$ in terms of matrices $A$ and $B$, acting in the parameters space $Y\equiv\{J,h\}$, namely:
\begin{align*}
\widetilde{Y}\equiv Y_1=A(\alpha)Y_0,& &\widetilde{C}\equiv C_1=B(\alpha)Y_0+C_0.
\end{align*}

Let's now iterate the renormalization procedure, choosing different parameters on different steps~(i.e. let's consider the general case). Then on the $n$-th step we will have
\begin{multline*}
Y_n=A(\alpha_n)A(\alpha_{n-1})\ldots A(\alpha_2)A(\alpha_1)Y_0,\\
\shoveleft{C_n=\left[B(\alpha_n)A(\alpha_{n-1})A(\alpha_{n-2})\ldots A(\alpha_2)A(\alpha_1)+\right.}\\
+B(\alpha_{n-1})A(\alpha_{n-2})A(\alpha_{n-3})\ldots A(\alpha_2)A(\alpha_1)+\ldots+\\
\left.+B(\alpha_2)A(\alpha_1)+B(\alpha_1)\right]Y_0+C_0.
\end{multline*}

But if we make an infinite number of steps as RSRG method requires for obtaining the estimate for the ground state energy, we'll get an expression with infinite number of parameters that will be impossible to minimize. That's why we suggest introducing new parameters only for several first steps and taking the same parameters for all next steps. Such an approach, on one hand, makes it possible to obtain a closed expression for the energy estimate; on the other hand, it gives more parameters to vary compared with the case when we take all parameters to be equal. Choosing different parameters only on the first three steps, for example, we will obtain the following expressions for the $n$-th ($n>3$) step:
\begin{multline*}
Y_n=A^{n-2}(\alpha_3)A(\alpha_2)A(\alpha_1)Y_0,\\
\shoveleft{C_n=\left[B(\alpha_3)\left(1+A(\alpha_3)+A^2(\alpha_3)+\ldots+A^{n-3}(\alpha_3)\right)
A(\alpha_2)A(\alpha_1)+\right.}\\
\left.+B(\alpha_2)A(\alpha_1)+B(\alpha_1)\right]Y_0+C_0.
\end{multline*}

Matrix $A$ possesses the following property: the absolute values of all its eigenvalues are less then one. So in the limit of $n\to\infty$, using the formula for a convergent geometrical progression, we will obtain
\begin{align*}
&Y_\infty=0,\\
&C_\infty=\left[B(\alpha_3)(1-A(\alpha_3))^{-1}A(\alpha_2)A(\alpha_1)
+B(\alpha_2)A(\alpha_1)+B(\alpha_1)\right]Y_0+C_0.
\end{align*}

$C_\infty$ now contains the information on the ground state energy. It is a closed equation with three parameters, which should be minimized to get an estimate for the energy.

\section{Calculation results}

The results of ground state energy calculations for different models, number of sites in a block, number of parameters and multi-steps renormalization are presented in tab.~1,~2,~3.

\begin{center}
Table 1. Calculation results for ITF model

\nopagebreak

\begin{tabular}{|c|c|c|c|}
\hline
\begin{tabular}{c} Number of Sites
\\
in a Block\end{tabular}
&
\begin{tabular}{c}
Error of the\\
Standard Method\\
\protect[Langari, 1998], \%
\end{tabular}
&
\begin{tabular}{c}
Number of\\
Parameters
\end{tabular}
&
\begin{tabular}{c}
Error of the \\
Modified Method
\end {tabular}
\\
\hline
2 &3.87&&
\\
\hline
2&&1&4.05
\\
\hline
2&&2&2.50
\\
\hline
2&&4&1.23
\\
\hline
2&&8&1.22
\\
\hline
\end{tabular}
\end{center}

\begin{center}
Table 2. Calculation results for $XX$-model

\nopagebreak

\begin{tabular}{|c|c|c|c|}
\hline
\begin{tabular}{c} Number of Sites
\\
in a Block\end{tabular}
&
\begin{tabular}{c}
Error of the\\
Standard Method\\
\protect[Langari, 1998], \%
\end{tabular}
&
\begin{tabular}{c}
Number of\\
Parameters
\end{tabular}
&
\begin{tabular}{c}
Error of the \\
Modified Method
\end {tabular}
\\
\hline
3&12.56&&\\
\hline
2&&8&9.56\\
\hline
$4=2\times2$&&16&7.42\\
\hline
4&&16&5.79\\
\hline
5&&54&4.86\\
\hline
$6=2\times3$&&21&8.20\\
\hline
$6=3\times2$&&24&2.20\\
\hline
$8=2\times4$&&21&7.39\\
\hline
$8=4\times2$&&30&4.08\\
\hline
$9=3\times3$&&28&6.31\\
\hline
$16=4\times4$&&72&5.49\\
\hline
\end{tabular}
\end{center}

\begin{center}
Table 3. Calculation results for isotropic Heisenberg model

\nopagebreak

\begin{tabular}{|c|c|c|c|}
\hline
\begin{tabular}{c} Number of Sites
\\
in a Block\end{tabular}
&
\begin{tabular}{c}
Error of the\\
Standard Method\\
\protect[Langari, 1998], \%
\end{tabular}
&
\begin{tabular}{c}
Number of\\
Parameters
\end{tabular}
&
\begin{tabular}{c}
Error of the \\
Modified Method
\end {tabular}
\\
\hline
3&13.25&&\\
\hline
2&&5&15.06\\
\hline
3&&16&12.42\\
\hline
$4=2\times2$&&12&12.19\\
\hline
4&&24&8.53\\
\hline
5&&54&7.66\\
\hline
$6=2\times3$&&21&14.52\\
\hline
$6=3\times2$&&24&6.82\\
\hline
$8=2\times4$&&24&6.82\\
\hline
$8=4\times2$&&30&6.43\\
\hline
$9=3\times3$&&28&10.55\\
\hline
$16=4\times4$&&72&6.42\\
\hline
\end{tabular}
\end{center}

In the tables above the column "Number of Sites in a Block"\, contains information on how many sites of the initial lattice were grouped in a block. If the value in the column contains a cross sign, e.g.~$6=3\times2$, it means, that a multi-steps blocking was used (in this example --- two-steps), and spins were grouped by three on the first step, and by two --- on the second.
The "Error of the Standard Method"\, column contains the calculation error for the energy in standard method compared with a known exact solution for each model. We should stress, that the developed method can be applied not only to exactly solvable models, but using them it's easy to check its effectiveness. The "Error of the Modified Method"\, column gives the error of the method described in this paper in percents compared to the exact result.

\section{Conclusions}

By analyzing the calculation results, it can be concluded, that:

\begin{itemize}
\item
the described method really improves an estimate for the ground state energy compared to the standard RSRG method;


\item
the suggested method also has its disadvantages. Thus, minimization of the estimate for the ground state energy doesn't always give an good estimate, the reason of which lies in the complicated structure of the function being minimized. And if the number of parameters is increased, the complexity of the function grows, and the minimization effectiveness decreases.

\end{itemize}

It should be once again stressed that the proposed method doesn't include any kind of hamiltonian diagonalization and that gives it an advantage both in accuracy and in time of producing the results compared to other methods (see Introduction).

Another advantage of it is that it always provides an estimate from above for the ground state energy, whereas other methods can't usually say if the true value is greater or lower than an obtained one. Adding an algorithm for providing an estimate from below will let us give a 100\% exact interval in which the true value lies, and we have some ideas on how to do that already.



\begin{thebibliography}{10}

\bibitem{burkhardt}
Burkhardt, T. W., van Leeuwen, J. M. J., Real-Space Renormalization, N.Y.: Springer-Verlag, 1982.

\bibitem{cardy}
Cardy, J., Scaling and Renormalization in Statistical Physics, Cambridge Univ. Press,1996.

\bibitem{continentino}
Continentino, M. A., Quantum Scaling in Many Body Systems, World Scientific Publishing, 2001.

\bibitem{drell}
Drell, S. D., Weinstein, M., Yankielowicz, S., Quantum Field Theories on a Lattice, Stanford, SLAC, 1977.

\bibitem{kadanoff}
Kadanoff, L. P., Scaling Laws for Ising Models Near $T_c$, {\it Physics}, 2, 263, 1966.

\bibitem{langari}
Langari, A., Phase Diagram of the Anti-ferromagnetic $XXZ$ Model in the Presence of an External Magnetic Field, {\it PRB}, 58, 14467-14475, 1998.

\bibitem{langariKarim}
Langari, A., Karimipor V., A Modified Quantum Renormalization Group for $XXZ$ Spin Chain, {\it Int. J. Mod. Phys. B}, 12, 2359, 1998.

\bibitem{malrieu}
Malrieu, J.-P., Guih\'ery N., Real-space Renormalization Group with Effective Interactions, {\it PRB}, 63, 085110, 2001.


\bibitem{martin-delgado}
Mart\'in-Delgado, M. A., Real-Space Renormalization Group Methods Applied to Quantum Lattice Hamiltonians, Proc. of the El Escorial Sum. Sch. on Strongly Correl. and Supercond. Syst., 1996.

\bibitem{ostlund}
\"Ostlund, S., Rommer, S., {\it PRL}, 75, 3537, 1995.

\bibitem{white}
White, S.R., Density Matrix Formulation for Quantum Renormalization Groups, {\it PRL}, 69, 2863, 1992.


\end{thebibliography}
\end{document}